\documentclass[12pt]{article}    

\newcommand{\be}{\begin{equation}}
\newcommand{\ee}{\end{equation}}
\newcommand{\bea}{\begin{eqnarray}}
\newcommand{\eea}{\end{eqnarray}}
\newcommand{\smallz}{{\scriptscriptstyle Z}} 
\newcommand{\smallw}{{\scriptscriptstyle W}} %
\newcommand{\smallh}{{\scriptscriptstyle H}} %
\newcommand{\mz}{M_\smallz}
\newcommand{\mw}{M_\smallw}
\newcommand{\mh}{M_\smallh}
\def \mt   {M_t}
\def \gev  {\mbox{ GeV}}
\def \seff {s^2_{eff}}
\def \ms   {\overline{\mbox{MS}}}
\def \cl   {\mbox{ 95\% C.L.}}

\begin{document}              

\begin{titlepage}
\begin{flushright}
        \small
        CERN-TH-97-197\\
        MPI-PhT/97-048\\
        NYU-TH/97-08-01\\
        hep-ph/9708311\\
        August 1997
\end{flushright}


\begin{center}
\vspace{1cm}
{\large\bf The Role of \boldmath{$\mw$} in Precision Studies of the 
                Standard Model}

\vspace{1cm}
\renewcommand{\thefootnote}{\fnsymbol{footnote}}
{\bf    G.~Degrassi$^a$\footnote
                {Permanent Address: Dipartimento di Fisica, Universit\`a
                  di Padova, Padova, Italy},
        P.~Gambino$^b$\footnote
                {E-mail address: gambino@mppmu.mpg.de},
        M.~Passera$^c$\footnote
                {E-mail address: massimo.passera@nyu.edu}, and
        A.~Sirlin$^b$\footnote
                {Permanent address: Dept. of Physics, New York 
                University, 4 Washington Place, New York, NY 10003, USA. 
                E-mail address: alberto.sirlin@nyu.edu}}
\setcounter{footnote}{0}
\vspace{.8cm}

{\it    $^a$ Theory Division, CERN CH-1211 Geneva 23, Switzerland \\
\vspace{3mm}
        $^b$ Max Planck Institut f\"{u}r Physik, W. Heisenberg Institut,\\
        F\"{o}hringer Ring 6, M\"{u}nchen, D-80805, Germany \\ 
\vspace{3mm}
        $^c$ Physics Department, Brookhaven National Laboratory, \\
        Upton, NY 11973-5000, USA}
\vspace{1.8cm}

{\large\bf Abstract}
\end{center}

\noindent
Recent calculations have significantly decreased the scheme and 
residual scale dependence of basic radiative corrections of the 
Standard Electroweak Model. This leads to a theoretically accurate 
prediction of the $W$-boson mass $\mw$, as well as a reduced upper bound 
for the Higgs boson mass $\mh$. The implications of a precise $\mw$
measurement on the $\mh$ estimate are emphasized.


\end{titlepage}

Two of the main objectives in current theoretical studies of the Standard 
Model (SM) are the improvement of the estimate of the Higgs boson mass
$\mh$ and its upper bound, and the accurate prediction of the $W$ boson
mass $\mw$. In this connection, theorists distinguish two types of errors:
parametric ones, which in principle can be reduced by improving experimental 
inputs, and theoretical uncertainties derived from the truncation of the 
perturbative series. The latter are usually estimated by comparing different
schemes of calculation that contain all the available theoretical 
information at a given order of accuracy. The difference between different
approaches is referred to as the scheme-dependence.

Because of its accuracy and its sensitivity to $\mh$, the effective 
electroweak mixing parameter $\sin^2 \!\theta^{lept}_{eff} \equiv \seff$,
determined at LEP and SLC, is of particular interest at present. 
Recent calculations of  $\seff$ and $\mw$ that incorporate reducible and 
irreducible contributions of $O(g^4 \mt^2/\mw^2)$ and use $\alpha$,
$G_\mu$, and $\mz$ as inputs \cite{DGV,DGS},
 examine three electroweak resummation 
approaches and two different ways of implementing the relevant QCD
corrections. One of the approaches ($\ms$) employs
$\hat{\alpha}(\mz)$ and $\sin^2 \!\hat{\theta}_{\smallw}(\mz) \equiv
\hat{s}^2$, the $\ms$ QED and electroweak mixing parameters
evaluated at the scale $\mu = \mz$, while the other two (OSI and OSII)
make use of the on-shell parameters $\alpha$ and 
$\sin^2 \!\theta_{\smallw} \equiv s^2 \equiv 1-\mw^2/\mz^2$. As expected, the
dependence on the electroweak scale $\mu$ cancels through 
$O(g^4 \mt^2/\mw^2)$. However, because complete $O(g^4)$ corrections
have not been evaluated, the $\ms$ and OSI formulations contain a
residual $O(g^4)$ scale dependence. On the other hand OSII is, by 
construction, strictly $\mu$-independent. It was shown in Ref.\cite{DGS} that
the incorporation of the irreducible  $O(g^4 \mt^2/\mw^2)$ corrections
sharply decreases the scheme dependence of the six calculations, to
the level of (4--5)$\times 10^{-5}$ in $\seff$ and 2--4 MeV in
$\mw$, depending on $\mh$. It is worth pointing out that such
variations are in rough accord with their expected order of
magnitude. In fact, we have \cite{FKS}
\be 
        \frac{\delta \seff}{\seff} \approx 
                \frac{\delta \hat{s}^2}{\hat{s}^2} \approx
                1.53 \; \delta \Delta\hat{r},
\ee
\be
        \frac{\delta \mw}{\mw} = -0.22 \;\delta \Delta r,
\ee
where $\delta \seff$ and $\delta \mw$ are the variations induced by
shifts $\delta \Delta\hat{r}$ and $\delta \Delta r$ of the basic
radiative corrections $\Delta\hat{r}$ \cite{si89} and $\Delta r$ 
\cite{si80}.
As the two-loop corrections that have not been included are not
enhanced by factors $(\mt^2/\mw^2)^n$ $(n=1,2)$, they may be expected
to be of $O(\hat{\alpha}/\pi \hat{s}^2)^2 \approx 10^{-4}$ in both 
$\Delta\hat{r}$ and $\Delta r$, implying 
$\delta \seff \approx 3.5\times 10^{-5}$ and $\delta \mw \approx 2$
MeV. The same argument suggests that the incorporated $O(g^4
\mt^2/\mw^2)$ corrections may be larger by a factor 4--5, which is 
roughly what has been observed at low $\mh$ values \cite{DGV,DGS}. 
As illustrated
in Fig.1, the irreducible $O(g^4\mt^2/\mw^2)$ corrections also sharply
reduce the residual electroweak scale dependence of the $\ms$ and OSI
approaches.

The main objective of this paper is to present simple analytic
formulae that reproduce to good accuracy the results of the new
calculations, and to show how they lead to a useful estimate of $\mh$
and its upper bound, and to a theoretically precise prediction of
$\mw$. The implications of an accurate $\mw$ measurement for the
$\mh$ estimate are also emphasized. 
The analytic formulae are of the form:
\bea
        \seff & = & (\seff)_0 + c_1 A_1 + c_2 A_2 - c_3 A_3 + c_4 A_4,  
                                                        \label{eq:s2}\\
        \mw   & = & \mw^0   - d_1 A_1 - d_5 A_1^2 - d_2 A_2 + d_3 A_3 
                - d_4 A_4,                              \label{eq:Mw}
\eea
where 
       $ A_1  \equiv  \ln(\mh / 100\gev)$,                 
       $ A_2  \equiv  [( \Delta \alpha)_h / 0.0280 -1]$,    
       \mbox{$ A_3  \equiv  [(\mt/175 \gev)^2 -1]$},  
       $ A_4  \equiv  [(\alpha_s(\mz)/0.118) -1]$,
$(\Delta \alpha)_h$ is the five-flavor hadronic contribution to the QED
vacuum-polarization function at $q^2=\mz^2$, and $(\seff)_0$ and $\mw^0$
are the theoretical results at the reference point 
$(\Delta \alpha)_h=0.0280$, $\mt = 175$ GeV, and $\alpha_s(\mz)=0.118$. 
The values of $(\seff)_0$, $\mw^0$, $c_i$ $(i=1-4)$, and 
$d_i$ $(i=1-5)$ for the three electroweak schemes of Ref.\cite{DGS}
and $\mz=91.1863$ GeV are given in Tables 1 and 2. For
brevity, we show the coefficients in the case of the 
$\mu_t$-parametrization, a procedure of implementing the QCD
corrections in which the pole top-quark mass $\mt$ is expressed in
terms of $\hat{m}_t (\mu_t) = \mu_t$, the $\ms$-parameter, leading
to sharply reduced QCD effects, and $\mu_t/\mt$ is evaluated by
optimization methods. In Ref.\cite{DGS} it was shown that in the three
electroweak schemes this method of implementing the QCD corrections
gives results very close to the direct use of $\mt$, an approach that
is frequently employed in the literature. 
In the range $75\gev \leq \mh \leq 350$ GeV, with the other parameters
within their $1-\sigma$ errors, Eq.~(\ref{eq:s2}) approximates the 
detailed calculations of Ref.\cite{DGS} with average absolute deviations of 
$\approx 4\times 10^{-6}$ and maximum absolute deviations of
$(1.1-1.3)\times 10^{-5}$, depending on the scheme;
Eq.~(\ref{eq:Mw}), which involves an additional parameter, shows
average absolute deviations of approximately 0.2 MeV and maximum absolute 
deviations of $(0.8-0.9)$ MeV. Outside the above range, the deviations 
increase reaching $(2.6-2.8)\times 10^{-5}$ and $(3.1-3.3)$ MeV at 
$\mh=600$ GeV.

We briefly discuss the estimation of $\mh$ from Eq.~(\ref{eq:s2}) and
the prediction of $\mw$ from Eq.~(\ref{eq:Mw}) using the direct 
experimental information on $\seff$, $(\Delta \alpha)_h$, $\mt$, and
$\alpha_s(\mz)$. From Eq.~(\ref{eq:s2}) we have
\be
        A_1 = A_1^c \pm \sigma_1,
\label{eq:A1}
\ee
\be
        A_1^c = \left[ \seff -(\seff)_0 -c_2 A_2 +c_3 A_3
                -c_4 A_4 \right]^c  \!\!/ c_1,
\label{eq:A1c}
\ee
\be
        \sigma_1 = \left[\sigma_s^2 +c_2^2 \sigma_2^2 
        +c_3^2 \sigma_3^2 +c_4^2 \sigma_4^2\right]^{\frac1{2}} \!\!/c_1, 
\label{eq:sig1}
\ee
where the superscript $c$ means that the central experimental values 
in $\seff$ and the $A_i$ $(i=1-4)$ are to be taken, and $\sigma_s$ and
$\sigma_i$ are the corresponding standard deviations. The predicted
central value of $\mw$ is obtained by inserting  $A_i^c$ in the r.h.s.
of Eq.~(\ref{eq:Mw}). Because $A_1$ is correlated with $A_i$ (cf. 
Eq.~(\ref{eq:A1c})) and Eq.~(\ref{eq:Mw}) contains a quadratic term in
$A_1$, the error analysis is slightly more involved in the $\mw$ case.
Defining 
        $\sigma_{\mw}^2 \equiv \overline{(\mw -\mw^c)^2}$
and taking into account the correlations one finds
\be
        \sigma_{\mw}^2 =  \sigma^2_1 \left[\hat{d_1}^2 +3 \,d_5^2  \,\sigma^2_1
                \right]+ \sum_{i=2}^4 \, d_i \sigma^2_i 
                \left[d_i -2\,\hat{d_1} c_i/c_1 \right ],
\label{eq:sigM}
\ee
where $\hat{d_1} = d_1 +2A_1^c \,d_5$. 
In the linear approximation ($d_5=0$) Eq.~(\ref{eq:sigM}) reduces to the 
simpler expression:
\be
        \sigma_{\mw}^2 = \left( d_1 \sigma_s /c_1 \right)^2 + \sum_{i=2}^4 
        \, \sigma_i^2 \left[d_i -d_1 c_i/c_1  \right]^2.
\label{eq:sigMapp}
\ee

We illustrate the application of these expressions using the first row
coefficients in Tables 1 and 2 ($\ms$ approach). Inserting the current
world averages
        $\seff = 0.23152 \pm 0.00023$ \cite{EWWG},
        $\mt   = 175.6   \pm 5.5$ GeV \cite{EWWG},
        $\alpha_s(\mz) = 0.118 \pm 0.005$ \cite{PDG}, and the evaluation 
        $(\Delta \alpha)_h = 0.02804 \pm 0.00065$ \cite{deltaalpha}
we find from Eqs.~(\ref{eq:A1}-\ref{eq:sig1})
\cite{footnote}: 
\be
        \ln \left(\mh/100 \right) = 0.029 \pm 0.709; \quad
        \ln \left(\mh/100 \right) < 1.195 ,
\label{eq:logMS}
\ee 
or, equivalently,
\be
        \mh = 103^{+106}_{-52} \gev; \quad \mh < 330 \gev ,
\ee
where  henceforth  the  
inequalities  represent  95\% C.L.  upper  bounds. From 
Eqs.~(\ref{eq:Mw},\ref{eq:sigM}) and (\ref{eq:logMS}) one obtains the 
prediction 
\be
        \mw = 80.384 \pm 0.033 \gev.
\label{eq:mwbound0}
\ee
We repeat this analysis for the other two schemes, with the results
listed in Table 3. We see that the three approaches give close
values. One way of combining them is to average the central values
of $\ln(\mh/100)$ and $\mw$ and expand the error to cover the range of
the three calculations. This gives
\be
        \ln \left(\mh/100 \right) = 0.000^{+0.738}_{-0.729}; \quad
        \ln \left(\mh/100 \right) < 1.214,
\ee
\be
        \mh = 100^{+109}_{-52} \gev; \quad \mh < 337 \gev ,
\label{eq:mhbound}
\ee
\be
        \mw = 80.384 \pm 0.034 \gev.
\label{eq:mwbound}
\ee

The dominant QCD contribution in these calculations is $\delta_{QCD}$,
the relevant correction in the evaluation of the electroweak parameter
$\Delta \rho$. For $\mt=175 \gev$, its theoretical error has been
estimated as $\pm 5.2 \times 10^{-3}$ \cite{sirqcd}. This induces errors of 
$\pm 1.8 \times 10^{-5}$ in $\seff$ and $\pm 3.1$ MeV in $\mw$, which
are of the same magnitude although somewhat larger than the
differences between the $\mu_t$ and $m_t$ parametrizations found in
Ref.\cite{DGS}. As there are additional QCD contributions,
 we may enlarge the
QCD theoretical error to $\pm 3\times 10^{-5}$ in $\seff$ and
$\pm 5$ MeV in $\mw$. An incremental uncertainty of $3\times
10^{-5}$ in $\seff$ shifts the $\mh$  bounds by $\approx\,6$\%, 
and Eqs.~(\ref{eq:mhbound},\ref{eq:mwbound}) become
\be
        \mh = 100^{+122}_{-54} \gev; \quad \mh < 357 \gev ,
\label{eq:mhbound1}
\ee
\be
        \mw = 80.384 \pm 0.039 \gev.
\label{eq:mwbound1}
\ee
Eq.~(\ref{eq:mwbound1}) is in good agreement with the current world
average, $
        \mw^{exp} = \left( 80.43 \pm 0.08 \right)
         $ GeV \cite{EWWG}.
Dividing Eq.~(\ref{eq:s2}) and Eq.~(\ref{eq:Mw}) by $(\seff)_0$ and
$\mw^0$, respectively, we see that, at equal levels of relative experimental
accuracy (which is, in fact, the current situation), $\seff$ is more
sensitive than $\mw$ by factors $\approx 2.7$ in $\ln(\mh/100)$, 6.6 in 
$(\Delta \alpha)_h$, 1.8 in $\mt^2$, 1.8 in $\alpha_s$. Thus, at
present, one of the main implications of Eq.~(\ref{eq:mwbound1}) 
and $\mw^{exp}$ is to
provide a sharp test of the SM, at the 0.1\% level, with very small
theoretical uncertainty.
There is, however, an important caveat in these considerations. The
current estimates of $\mh$ depend crucially on the world average 
$\seff = 0.23152 \pm 0.00023$, and this follows from a combination of
experimental results that are not in good harmony. For example, the
current $\chi^2$/d.o.f. in the determination of $\seff$ from LEP + SLC
asymmetries is $12.5/6$, with a C.L. of 5.2\%. As a rough and perhaps
extreme illustration of this situation, we note that employing only
the LEP average $\seff = 0.23196 \pm 0.00028$ one obtains a
95\% C.L. upper bound for $\mh$ larger than 800 GeV, while 
using the SLD value $\seff = 0.23055 \pm 0.00041$ alone the corresponding
upper bound is $\approx 80$ GeV. It is clear that the $\mh$ upper
bound depends in a very sensitive manner on the precise central value
and error of $\sin^2 \! \theta^{lept}_{eff}$: using the world average
one obtains an interesting constraint; however, the LEP value alone
leads to a very loose constraint and the SLD value alone is barely
compatible with the lower limit on $\mh$ from direct searches! 
If the $\seff$ error is increased
by a scaling factor $S=[\chi^2/(N-1)]^{1/2}$ according to Particle data group
prescription \cite{PDG}, we have $\seff = 0.23152 \pm 0.00033$. Combining the 
results of the three schemes as before one finds
\be
        \mh = 100^{+153}_{-60} \gev; \quad \mh < 443 \gev ,
\label{eq:mhbound2}
\ee
where we have included the QCD uncertainty. Although this scaling method is
not generally employed in current analyses of the electroweak data
(an exception is Ref. \cite{gurtu}),
it provides a more conservative and perhaps more realistic estimate of 
$\mh$.

This state of affairs strongly suggests the desirability of obtaining
constraints on $\mh$ derived from future precise measurements of $\mw$.
Using $\mw$ as an input, we see from Eq.~(\ref{eq:Mw}) that 
$ y \equiv d_1 A_1 + d_5 A_1^2$ is normally distributed about
\be
        \ y^c=  \mw^0 -\mw^c   - d_2 A_2 + d_3 A_3 
                - d_4 A_4,                    
          \label{eq:Mwy}
\ee
with standard deviation 
\be
        \sigma_y = \left[ \sigma_{\mw}^2 +
          \sum_{i=2}^4 d_i^2 \sigma_i^2 \right]^{\frac1{2}} . 
\label{eq:sigy}
\ee
As an illustration, we assume future measurements of $\mw$ and $\mt$ with
$\sigma_{\mw} = 35$ MeV, $\sigma_{\mt} = 3 \,\gev$, without changes in 
$(\Delta \alpha)_h$, $\alpha_s (\mz)$ and $\mt^c$.
To compare the sensitivity of such a measurement with the current $\seff$ 
determination, we assume $\mw^c = 80.384 \, \gev$, i.e.~the value predicted in
Eq.~(\ref{eq:mwbound0}) and Eq.~(\ref{eq:mwbound}). Using the $\ms$
scheme (first row of Table 2), Eqs.~(\ref{eq:Mwy},\ref{eq:sigy}) lead to
$ y = 0.0017 \pm 0.0417$, $y< 0.0703$, which correspond to
\be
        \mh = 103^{+95}_{-57} \gev; \quad \mh < 288 \gev .
\label{eq:mhbound3}
\ee
The QCD uncertainty $\Delta\, \mw = \pm 5$ MeV increases the $\cl$ upper bound 
to $308 \gev$. 

Comparison of Eqs.~(\ref{eq:mhbound}) and (\ref{eq:mhbound3}) shows that an
$\mw$ determination of $\mh$ with $\sigma_{\mw} = 0.035\, \gev$ and
$\sigma_{\mt} = 3\, \gev$ would be somewhat more restrictive than the current
$\seff$ estimate. This scenario is consistent with the expectation
$\sigma_{\mw} \approx 40\!-\!50$ MeV, $\sigma_{\mt} \approx 3 \, \gev$ in 
Tevatron
Run 2 and $\sigma_{\mw} \approx 20\!-\!30$ MeV, $\sigma_{\mt} \approx 
1\!-\!1.5$
GeV  in Run 3 \cite{tevatron}. One also expects 
increased accuracy in the $\mw$ measurements at LEPII. An important element
in the $\mh$ estimate based on $\mw$ is the insensitivity to 
$(\Delta \alpha)_h$, a parameter whose future accuracy is not clear at 
present. On the contrary, in the $\mh$ estimate based on $\seff$, 
$(\Delta \alpha)_h$ is responsible for one of the largest uncertainties.
It is also worth pointing out that the current world average central
value $\mw = 80.43 \, \gev$ favors smaller $\mh$ values than 
Eqs.~(\ref{eq:mhbound},\ref{eq:mhbound2}). For example, assuming a
future measurement $\mw = 80.430 \pm 0.035\, \gev$ and $\sigma_{\mt} = 3 \, 
\gev$
we would get in the $\ms$ scheme $\mh < 149 \, \gev\, (\cl)$ which would be 
a very interesting constraint.

Next, we consider the simultaneous use of $(\seff)^{exp}$ and $\mw^{exp}$
in the $\mh$ estimate. Because of the correlations and quadratic form of 
Eq.~(\ref{eq:Mw}), this is most easily done with a numerical 
$\chi^2$-analysis employing the theoretical expressions of 
Eqs.~(\ref{eq:s2},\ref{eq:Mw}). Using the $\ms$ scheme,
Table 4 gives the $\mh$ values and the $\cl$
upper bounds for the current experimental inputs and for a future scenario
with $\sigma_{\mw} = 35$ MeV and $\sigma_{\mt} = 3\, \gev$. In both cases
we employ the conventional and scaled versions of the $\seff$ uncertainty.
We see that the constraints in the future scenarios are somewhat less 
restrictive than when we consider 
 $\mw$ alone.  This is due to the fact that the present $\mw^c$ leads to a 
lower $\mh$ value than the one derived from $(\seff)^c$.

Finally, it is important to note that the incorporation of the 
$O(g^4 \mt^2/\mw^2)$ terms, as implemented in Refs.~\cite{DGV,DGS} leads,
for equal inputs, to significantly lower $\mh$ estimates than those 
obtained in conventional calculations which do not include such 
contributions. As an illustration, we consider a recent fit to the 
electroweak data and $\mt^{exp}$ \cite{EWWG} which yields 
$\sin^2 \!\theta^{lept}_{eff} = 0.23152 \pm 0.00022$,
$\mt = 173.1 \pm 5.4$ GeV,
$\alpha(\mz)^{-1} = 128.898 \pm 0.090$ 
(corresponding to $(\Delta \alpha)_h = 0.02803 \pm 0.00065$),
$\alpha_s(\mz) = 0.120 \pm 0.003$, and an indirect determination 
$\mh = 115^{+116}_{-66}$ GeV. For the same input data for 
$\sin^2 \!\theta^{lept}_{eff}$, $\mt$, $(\Delta \alpha)_h$,
$\alpha_s(\mz)$, Eq.~(\ref{eq:s2}) in the $\ms$ scheme leads to 
$\mh = 88^{+87}_{-44}$ GeV (the main difference with Eq.~(\ref{eq:logMS})
is due to the fact that the fit to the electroweak data lowers the value of 
$\mt$). We see that the central value and $1\sigma$ upper 
bound estimated in the analysis of the fit are about 30\% larger than the 
value derived in the $\ms$ scheme from Eq.~(\ref{eq:s2}). If we use the 
same inputs in the OSI and OSII schemes, average the $\ln(\mh/100)$ 
results as before and include the estimate of the QCD error, we find 
$\mh = 85^{+100}_{-46}$ GeV, $\mh < 295$ GeV (95\% C.L.). 
The fact that the latter is significantly smaller than the 95\% C.L. 
upper bound 420 GeV reported in Ref.~\cite{EWWG}, which includes an estimate
of all theoretical errors, is due not only to the $\approx 30\%$ effect 
explained before, but also to the fact that the scheme dependence is 
considerably larger when the $O(g^4 \mt^2/\mw^2)$ terms are not included.

\subsection*{Acknowledgments}
The authors \, are \,indebted \,to \,G.~Altarelli, \,W.A.~Bardeen, \,
S.~Fanchiotti, \, and \,
W.J.~Marciano for  very useful discussions. 
The work of M.P.~ and A.S.~was  supported 
in part by the U.S. Department of Energy under 
Contract No. DE-AC02-76-CH00016 and by the NSF Grant PHY-9722083,    
respectively.


\newpage

\renewcommand{\arraystretch}{1.15}
\begin{table}
\begin{center}
\begin{tabular}{|c||c|c|c|c|c|} 
\hline 
 Scheme & $(\seff)_0$  & $10^4 c_1$  & $10^3 c_2$  & $10^3 c_3$ & 
        $10^4 c_4$  \\ \hline \hline 
$\ms$          &0.231510 &  5.23 & 9.86 & 2.78 & 4.5    \\ \hline 
OSI            &0.231524 &  5.19 & 9.86 & 2.77 & 4.5    \\ \hline
OSII           &0.231540 &  5.26 & 9.86 & 2.68 & 4.4    \\ \hline
\end{tabular} 
\caption{Values of $(\seff)_0$ and $c_i$ $(i=1-4)$ in Eq.~(3)
        for three electroweak schemes that incorporate $O(g^4\mt^2/\mw^2)$
        corrections in the $\mu_t$-parametrization of QCD corrections 
        [2].
}
\end{center} 
\end{table} 
\begin{table}
\begin{center}
\begin{tabular}{|c||c|c|c|c|c|c|} 
\hline 
 Scheme & $\mw^0$  & $10^2 d_1$  & $10 \,d_2$  & $10\, d_3$ & 
         $10^2 d_4$  & $10^3 d_5$ \\ \hline \hline 
$\ms$          & 80.3827 &  5.79 & 5.17 & 5.43 & 8.5 & 8.0      \\ \hline 
OSI            & 80.3807 &  5.73 & 5.18 & 5.41 & 8.5 & 8.0      \\ \hline 
OSII           & 80.3805 &  5.81 & 5.18 & 5.37 & 8.5 & 7.8      \\ \hline 
\end{tabular} 
\caption{Values of $\mw^0$ and $d_i$ $(i=1-5)$ in Eq.~(4),
        in GeV, for the same electroweak schemes as in Table 1.}
\end{center} 
\end{table} 
\begin{table}
\begin{center}
\begin{tabular}{|c||c|c|c|} 
\hline 
 Scheme & $\ln(\mh /100)$       & $\mh$ (GeV)              & $\mw$ (GeV)    
                                                        \\ \hline \hline 
$\ms$   & $0.029  \pm 0.709$  & $103^{+106}_{-52}$ & $80.384 \pm 0.033$
                                                        \\ \hline 
OSI     & $0.002 \pm 0.714$  & $100^{+104}_{-51}$  & $80.384 \pm 0.033$
                                                        \\ \hline 
OSII    & $-0.030 \pm 0.699$  & $97^{+98}_{-49}$   & $80.385 \pm 0.033$
                                                        \\ \hline 
\end{tabular} 
\caption{Values of $\mh$ and $\mw$ obtained from the current world averages
        of $\seff$, $\mt$, $\hat{\alpha}_s$, and the evaluation of 
        $(\Delta \alpha)_h$ [8], on the basis of 
        Eqs.~(3,4) and Tables 1,2.}
\end{center} 
\end{table}
\begin{table}
\begin{center}
\begin{tabular}{|c||c|c|} 
\hline 
 scenario    &  $\mh$ (GeV)  & upper bound     \\ \hline \hline 
current conv.     &   96$^{+103}_{-49}$    &305             
                                                        \\ \hline 
current scaled    &  90$^{+116}_{-51}$    &332          
                                                    \\ \hline 
future conv.     &  76$^{+60}_{-35}$ &     188      
                                                        \\ \hline       
future scaled    & 68$^{+61}_{-34}$   &182 
                                                        \\ \hline 
\end{tabular} 
\caption{Values and 95\% C.L. upper bounds of $\mh$  obtained 
in the $\ms$ scheme
combining $\seff$ and $\mw$ data and expressed in GeV.
 The current scenario  involves the present 
experimental values, the future scenario assumes $\sigma_{\mw}=35$MeV and 
$\sigma_{\mt}=3$GeV with unchanged central values.
 The error on $\seff$ is  taken to be 2.3$\times10^{-4}$ (conventional) or 
 3.3$\times10^{-4}$ (scaled). 
The theoretical   uncertainty
due to QCD corrections is included as a systematic error. 
}
\end{center} 
\end{table}


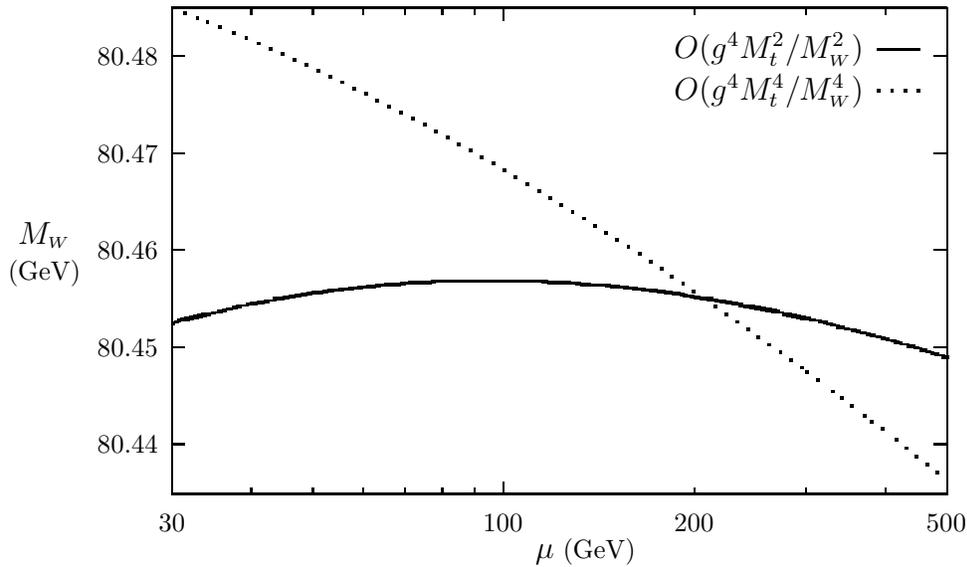
\begin{figure}

\setlength{\unitlength}{0.240900pt}
\ifx\plotpoint\undefined\newsavebox{\plotpoint}\fi
\sbox{\plotpoint}{\rule[-0.200pt]{0.400pt}{0.400pt}}%
\begin{picture}(1500,900)(0,0)
\font\gnuplot=cmr10 at 10pt
\gnuplot
\sbox{\plotpoint}{\rule[-0.200pt]{0.400pt}{0.400pt}}%
\put(220.0,191.0){\rule[-0.200pt]{4.818pt}{0.400pt}}
\put(198,191){\makebox(0,0)[r]{80.44}}
\put(1416.0,191.0){\rule[-0.200pt]{4.818pt}{0.400pt}}
\put(220.0,343.0){\rule[-0.200pt]{4.818pt}{0.400pt}}
\put(198,343){\makebox(0,0)[r]{80.45}}
\put(1416.0,343.0){\rule[-0.200pt]{4.818pt}{0.400pt}}
\put(220.0,496.0){\rule[-0.200pt]{4.818pt}{0.400pt}}
\put(198,496){\makebox(0,0)[r]{80.46}}
\put(1416.0,496.0){\rule[-0.200pt]{4.818pt}{0.400pt}}
\put(220.0,648.0){\rule[-0.200pt]{4.818pt}{0.400pt}}
\put(198,648){\makebox(0,0)[r]{80.47}}
\put(1416.0,648.0){\rule[-0.200pt]{4.818pt}{0.400pt}}
\put(220.0,801.0){\rule[-0.200pt]{4.818pt}{0.400pt}}
\put(198,801){\makebox(0,0)[r]{80.48}}
\put(1416.0,801.0){\rule[-0.200pt]{4.818pt}{0.400pt}}
\put(220.0,113.0){\rule[-0.200pt]{0.400pt}{2.409pt}}
\put(220.0,867.0){\rule[-0.200pt]{0.400pt}{2.409pt}}
\put(344.0,113.0){\rule[-0.200pt]{0.400pt}{2.409pt}}
\put(344.0,867.0){\rule[-0.200pt]{0.400pt}{2.409pt}}
\put(441.0,113.0){\rule[-0.200pt]{0.400pt}{2.409pt}}
\put(441.0,867.0){\rule[-0.200pt]{0.400pt}{2.409pt}}
\put(520.0,113.0){\rule[-0.200pt]{0.400pt}{2.409pt}}
\put(520.0,867.0){\rule[-0.200pt]{0.400pt}{2.409pt}}
\put(586.0,113.0){\rule[-0.200pt]{0.400pt}{2.409pt}}
\put(586.0,867.0){\rule[-0.200pt]{0.400pt}{2.409pt}}
\put(644.0,113.0){\rule[-0.200pt]{0.400pt}{2.409pt}}
\put(644.0,867.0){\rule[-0.200pt]{0.400pt}{2.409pt}}
\put(695.0,113.0){\rule[-0.200pt]{0.400pt}{2.409pt}}
\put(695.0,867.0){\rule[-0.200pt]{0.400pt}{2.409pt}}
\put(740.0,113.0){\rule[-0.200pt]{0.400pt}{4.818pt}}
\put(220,68){\makebox(0,0){30}}
\put(740,68){\makebox(0,0){100}}
\put(1040,68){\makebox(0,0){200}}
\put(1436,68){\makebox(0,0){500}}
\put(740.0,857.0){\rule[-0.200pt]{0.400pt}{4.818pt}}
\put(1040.0,113.0){\rule[-0.200pt]{0.400pt}{2.409pt}}
\put(1040.0,867.0){\rule[-0.200pt]{0.400pt}{2.409pt}}
\put(1215.0,113.0){\rule[-0.200pt]{0.400pt}{2.409pt}}
\put(1215.0,867.0){\rule[-0.200pt]{0.400pt}{2.409pt}}
\put(1340.0,113.0){\rule[-0.200pt]{0.400pt}{2.409pt}}
\put(1340.0,867.0){\rule[-0.200pt]{0.400pt}{2.409pt}}
\put(1436.0,113.0){\rule[-0.200pt]{0.400pt}{2.409pt}}
\put(1436.0,867.0){\rule[-0.200pt]{0.400pt}{2.409pt}}
\put(220.0,113.0){\rule[-0.200pt]{292.934pt}{0.400pt}}
\put(1436.0,113.0){\rule[-0.200pt]{0.400pt}{184.048pt}}
\put(220.0,877.0){\rule[-0.200pt]{293.334pt}{0.400pt}}
\put(20,520){\makebox(0,0){$\mw$}}
\put(20,460){\makebox(0,0){(GeV)}}
\put(870,23){\makebox(0,0){$\mu$ (GeV)}}
\put(220.0,113.0){\rule[-0.200pt]{0.400pt}{184.048pt}}
\sbox{\plotpoint}{\rule[-0.400pt]{0.800pt}{0.800pt}}%
\put(1306,812){\makebox(0,0)[r]{$O (g^4 M_t^2/\mw^2)$}}
\put(1328.0,812.0){\rule[-0.400pt]{15.899pt}{0.800pt}}
\put(220,381){\usebox{\plotpoint}}
\multiput(220.00,382.38)(1.936,0.560){3}{\rule{2.440pt}{0.135pt}}
\multiput(220.00,379.34)(8.936,5.000){2}{\rule{1.220pt}{0.800pt}}
\put(234,385.84){\rule{3.373pt}{0.800pt}}
\multiput(234.00,384.34)(7.000,3.000){2}{\rule{1.686pt}{0.800pt}}
\put(248,388.84){\rule{3.132pt}{0.800pt}}
\multiput(248.00,387.34)(6.500,3.000){2}{\rule{1.566pt}{0.800pt}}
\put(261,391.84){\rule{3.132pt}{0.800pt}}
\multiput(261.00,390.34)(6.500,3.000){2}{\rule{1.566pt}{0.800pt}}
\multiput(274.00,396.39)(2.583,0.536){5}{\rule{3.533pt}{0.129pt}}
\multiput(274.00,393.34)(17.666,6.000){2}{\rule{1.767pt}{0.800pt}}
\multiput(299.00,402.39)(2.360,0.536){5}{\rule{3.267pt}{0.129pt}}
\multiput(299.00,399.34)(16.220,6.000){2}{\rule{1.633pt}{0.800pt}}
\multiput(322.00,408.38)(3.279,0.560){3}{\rule{3.720pt}{0.135pt}}
\multiput(322.00,405.34)(14.279,5.000){2}{\rule{1.860pt}{0.800pt}}
\put(344,411.84){\rule{5.059pt}{0.800pt}}
\multiput(344.00,410.34)(10.500,3.000){2}{\rule{2.529pt}{0.800pt}}
\multiput(365.00,416.38)(3.111,0.560){3}{\rule{3.560pt}{0.135pt}}
\multiput(365.00,413.34)(13.611,5.000){2}{\rule{1.780pt}{0.800pt}}
\put(386,419.84){\rule{4.577pt}{0.800pt}}
\multiput(386.00,418.34)(9.500,3.000){2}{\rule{2.289pt}{0.800pt}}
\put(405,422.84){\rule{4.336pt}{0.800pt}}
\multiput(405.00,421.34)(9.000,3.000){2}{\rule{2.168pt}{0.800pt}}
\put(423,425.84){\rule{4.336pt}{0.800pt}}
\multiput(423.00,424.34)(9.000,3.000){2}{\rule{2.168pt}{0.800pt}}
\put(441,427.84){\rule{4.095pt}{0.800pt}}
\multiput(441.00,427.34)(8.500,1.000){2}{\rule{2.048pt}{0.800pt}}
\put(458,429.84){\rule{3.854pt}{0.800pt}}
\multiput(458.00,428.34)(8.000,3.000){2}{\rule{1.927pt}{0.800pt}}
\put(474,432.34){\rule{3.854pt}{0.800pt}}
\multiput(474.00,431.34)(8.000,2.000){2}{\rule{1.927pt}{0.800pt}}
\put(490,433.84){\rule{3.614pt}{0.800pt}}
\multiput(490.00,433.34)(7.500,1.000){2}{\rule{1.807pt}{0.800pt}}
\put(505,435.34){\rule{3.614pt}{0.800pt}}
\multiput(505.00,434.34)(7.500,2.000){2}{\rule{1.807pt}{0.800pt}}
\put(520,436.84){\rule{3.373pt}{0.800pt}}
\multiput(520.00,436.34)(7.000,1.000){2}{\rule{1.686pt}{0.800pt}}
\put(534,438.34){\rule{3.132pt}{0.800pt}}
\multiput(534.00,437.34)(6.500,2.000){2}{\rule{1.566pt}{0.800pt}}
\put(547,439.84){\rule{3.373pt}{0.800pt}}
\multiput(547.00,439.34)(7.000,1.000){2}{\rule{1.686pt}{0.800pt}}
\put(574,441.34){\rule{2.891pt}{0.800pt}}
\multiput(574.00,440.34)(6.000,2.000){2}{\rule{1.445pt}{0.800pt}}
\put(586,442.84){\rule{2.891pt}{0.800pt}}
\multiput(586.00,442.34)(6.000,1.000){2}{\rule{1.445pt}{0.800pt}}
\put(561.0,442.0){\rule[-0.400pt]{3.132pt}{0.800pt}}
\put(622,444.34){\rule{2.650pt}{0.800pt}}
\multiput(622.00,443.34)(5.500,2.000){2}{\rule{1.325pt}{0.800pt}}
\put(598.0,445.0){\rule[-0.400pt]{5.782pt}{0.800pt}}
\put(660,445.84){\rule{3.614pt}{0.800pt}}
\multiput(660.00,445.34)(7.500,1.000){2}{\rule{1.807pt}{0.800pt}}
\put(633.0,447.0){\rule[-0.400pt]{6.504pt}{0.800pt}}
\put(761,445.84){\rule{5.059pt}{0.800pt}}
\multiput(761.00,446.34)(10.500,-1.000){2}{\rule{2.529pt}{0.800pt}}
\put(675.0,448.0){\rule[-0.400pt]{20.717pt}{0.800pt}}
\put(801,444.34){\rule{4.336pt}{0.800pt}}
\multiput(801.00,445.34)(9.000,-2.000){2}{\rule{2.168pt}{0.800pt}}
\put(782.0,447.0){\rule[-0.400pt]{4.577pt}{0.800pt}}
\put(837,442.84){\rule{4.095pt}{0.800pt}}
\multiput(837.00,443.34)(8.500,-1.000){2}{\rule{2.048pt}{0.800pt}}
\put(854,441.34){\rule{3.854pt}{0.800pt}}
\multiput(854.00,442.34)(8.000,-2.000){2}{\rule{1.927pt}{0.800pt}}
\put(870,439.84){\rule{3.854pt}{0.800pt}}
\multiput(870.00,440.34)(8.000,-1.000){2}{\rule{1.927pt}{0.800pt}}
\put(819.0,445.0){\rule[-0.400pt]{4.336pt}{0.800pt}}
\put(901,438.34){\rule{3.614pt}{0.800pt}}
\multiput(901.00,439.34)(7.500,-2.000){2}{\rule{1.807pt}{0.800pt}}
\put(916,436.84){\rule{3.373pt}{0.800pt}}
\multiput(916.00,437.34)(7.000,-1.000){2}{\rule{1.686pt}{0.800pt}}
\put(930,435.34){\rule{3.373pt}{0.800pt}}
\multiput(930.00,436.34)(7.000,-2.000){2}{\rule{1.686pt}{0.800pt}}
\put(944,433.84){\rule{3.132pt}{0.800pt}}
\multiput(944.00,434.34)(6.500,-1.000){2}{\rule{1.566pt}{0.800pt}}
\put(957,432.34){\rule{3.132pt}{0.800pt}}
\multiput(957.00,433.34)(6.500,-2.000){2}{\rule{1.566pt}{0.800pt}}
\put(970,430.84){\rule{2.891pt}{0.800pt}}
\multiput(970.00,431.34)(6.000,-1.000){2}{\rule{1.445pt}{0.800pt}}
\put(982,429.34){\rule{2.891pt}{0.800pt}}
\multiput(982.00,430.34)(6.000,-2.000){2}{\rule{1.445pt}{0.800pt}}
\put(994,427.84){\rule{2.891pt}{0.800pt}}
\multiput(994.00,428.34)(6.000,-1.000){2}{\rule{1.445pt}{0.800pt}}
\put(1006,426.34){\rule{2.891pt}{0.800pt}}
\multiput(1006.00,427.34)(6.000,-2.000){2}{\rule{1.445pt}{0.800pt}}
\put(1018,423.84){\rule{2.650pt}{0.800pt}}
\multiput(1018.00,425.34)(5.500,-3.000){2}{\rule{1.325pt}{0.800pt}}
\put(1029,421.84){\rule{2.650pt}{0.800pt}}
\multiput(1029.00,422.34)(5.500,-1.000){2}{\rule{1.325pt}{0.800pt}}
\put(1040,419.84){\rule{5.059pt}{0.800pt}}
\multiput(1040.00,421.34)(10.500,-3.000){2}{\rule{2.529pt}{0.800pt}}
\put(1061,416.34){\rule{4.200pt}{0.800pt}}
\multiput(1061.00,418.34)(11.283,-4.000){2}{\rule{2.100pt}{0.800pt}}
\put(1081,412.84){\rule{4.577pt}{0.800pt}}
\multiput(1081.00,414.34)(9.500,-3.000){2}{\rule{2.289pt}{0.800pt}}
\put(1100,409.34){\rule{4.000pt}{0.800pt}}
\multiput(1100.00,411.34)(10.698,-4.000){2}{\rule{2.000pt}{0.800pt}}
\put(1119,405.84){\rule{4.095pt}{0.800pt}}
\multiput(1119.00,407.34)(8.500,-3.000){2}{\rule{2.048pt}{0.800pt}}
\put(1136,402.84){\rule{4.095pt}{0.800pt}}
\multiput(1136.00,404.34)(8.500,-3.000){2}{\rule{2.048pt}{0.800pt}}
\put(1153,399.84){\rule{4.095pt}{0.800pt}}
\multiput(1153.00,401.34)(8.500,-3.000){2}{\rule{2.048pt}{0.800pt}}
\multiput(1170.00,398.06)(2.104,-0.560){3}{\rule{2.600pt}{0.135pt}}
\multiput(1170.00,398.34)(9.604,-5.000){2}{\rule{1.300pt}{0.800pt}}
\put(1185,391.84){\rule{3.854pt}{0.800pt}}
\multiput(1185.00,393.34)(8.000,-3.000){2}{\rule{1.927pt}{0.800pt}}
\put(1201,388.84){\rule{3.373pt}{0.800pt}}
\multiput(1201.00,390.34)(7.000,-3.000){2}{\rule{1.686pt}{0.800pt}}
\multiput(1215.00,387.07)(2.918,-0.536){5}{\rule{3.933pt}{0.129pt}}
\multiput(1215.00,387.34)(19.836,-6.000){2}{\rule{1.967pt}{0.800pt}}
\multiput(1243.00,381.08)(1.797,-0.520){9}{\rule{2.800pt}{0.125pt}}
\multiput(1243.00,381.34)(20.188,-8.000){2}{\rule{1.400pt}{0.800pt}}
\multiput(1269.00,373.07)(2.583,-0.536){5}{\rule{3.533pt}{0.129pt}}
\multiput(1269.00,373.34)(17.666,-6.000){2}{\rule{1.767pt}{0.800pt}}
\multiput(1294.00,367.07)(2.360,-0.536){5}{\rule{3.267pt}{0.129pt}}
\multiput(1294.00,367.34)(16.220,-6.000){2}{\rule{1.633pt}{0.800pt}}
\multiput(1317.00,361.07)(2.360,-0.536){5}{\rule{3.267pt}{0.129pt}}
\multiput(1317.00,361.34)(16.220,-6.000){2}{\rule{1.633pt}{0.800pt}}
\multiput(1340.00,355.07)(2.137,-0.536){5}{\rule{3.000pt}{0.129pt}}
\multiput(1340.00,355.34)(14.773,-6.000){2}{\rule{1.500pt}{0.800pt}}
\multiput(1361.00,349.07)(2.025,-0.536){5}{\rule{2.867pt}{0.129pt}}
\multiput(1361.00,349.34)(14.050,-6.000){2}{\rule{1.433pt}{0.800pt}}
\multiput(1381.00,343.07)(1.913,-0.536){5}{\rule{2.733pt}{0.129pt}}
\multiput(1381.00,343.34)(13.327,-6.000){2}{\rule{1.367pt}{0.800pt}}
\multiput(1400.00,337.07)(1.802,-0.536){5}{\rule{2.600pt}{0.129pt}}
\multiput(1400.00,337.34)(12.604,-6.000){2}{\rule{1.300pt}{0.800pt}}
\multiput(1418.00,331.06)(2.607,-0.560){3}{\rule{3.080pt}{0.135pt}}
\multiput(1418.00,331.34)(11.607,-5.000){2}{\rule{1.540pt}{0.800pt}}
\put(886.0,441.0){\rule[-0.400pt]{3.613pt}{0.800pt}}
\put(1436,328){\usebox{\plotpoint}}
\sbox{\plotpoint}{\rule[-0.500pt]{1.000pt}{1.000pt}}%
\put(1306,747){\makebox(0,0)[r]{$O (g^4 M_t^4/\mw^4)$}}
\multiput(1328,747)(20.756,0.000){4}{\usebox{\plotpoint}}
\put(239.08,868.82){\usebox{\plotpoint}}
\put(258.31,861.03){\usebox{\plotpoint}}
\multiput(261,860)(18.845,-8.698){0}{\usebox{\plotpoint}}
\multiput(274,854)(19.529,-7.030){2}{\usebox{\plotpoint}}
\put(315.69,837.02){\usebox{\plotpoint}}
\put(334.73,828.79){\usebox{\plotpoint}}
\put(353.88,820.77){\usebox{\plotpoint}}
\put(372.95,812.59){\usebox{\plotpoint}}
\put(391.93,804.19){\usebox{\plotpoint}}
\put(410.75,795.44){\usebox{\plotpoint}}
\put(429.57,786.71){\usebox{\plotpoint}}
\put(448.22,777.60){\usebox{\plotpoint}}
\put(467.11,769.01){\usebox{\plotpoint}}
\put(485.84,760.08){\usebox{\plotpoint}}
\put(504.21,750.42){\usebox{\plotpoint}}
\multiput(505,750)(18.808,-8.777){0}{\usebox{\plotpoint}}
\put(522.87,741.36){\usebox{\plotpoint}}
\put(540.99,731.24){\usebox{\plotpoint}}
\put(559.46,721.77){\usebox{\plotpoint}}
\multiput(561,721)(18.275,-9.840){0}{\usebox{\plotpoint}}
\put(577.82,712.09){\usebox{\plotpoint}}
\put(596.38,702.81){\usebox{\plotpoint}}
\multiput(598,702)(18.564,-9.282){0}{\usebox{\plotpoint}}
\put(614.78,693.21){\usebox{\plotpoint}}
\put(632.88,683.07){\usebox{\plotpoint}}
\multiput(633,683)(18.221,-9.939){0}{\usebox{\plotpoint}}
\put(651.05,673.03){\usebox{\plotpoint}}
\put(668.99,662.60){\usebox{\plotpoint}}
\put(687.13,652.53){\usebox{\plotpoint}}
\multiput(690,651)(17.459,-11.224){0}{\usebox{\plotpoint}}
\put(704.77,641.61){\usebox{\plotpoint}}
\put(723.18,632.04){\usebox{\plotpoint}}
\multiput(732,627)(17.601,-11.000){0}{\usebox{\plotpoint}}
\multiput(740,622)(17.648,-10.925){2}{\usebox{\plotpoint}}
\put(776.28,599.54){\usebox{\plotpoint}}
\put(794.42,589.46){\usebox{\plotpoint}}
\put(812.08,578.61){\usebox{\plotpoint}}
\put(829.61,567.51){\usebox{\plotpoint}}
\put(847.16,556.43){\usebox{\plotpoint}}
\put(864.99,545.82){\usebox{\plotpoint}}
\put(882.36,534.50){\usebox{\plotpoint}}
\put(900.01,523.59){\usebox{\plotpoint}}
\multiput(901,523)(17.270,-11.513){0}{\usebox{\plotpoint}}
\put(917.28,512.08){\usebox{\plotpoint}}
\put(934.59,500.71){\usebox{\plotpoint}}
\put(952.41,490.18){\usebox{\plotpoint}}
\put(969.03,477.75){\usebox{\plotpoint}}
\multiput(970,477)(17.928,-10.458){0}{\usebox{\plotpoint}}
\put(986.51,466.62){\usebox{\plotpoint}}
\put(1003.48,454.68){\usebox{\plotpoint}}
\multiput(1006,453)(17.270,-11.513){0}{\usebox{\plotpoint}}
\put(1020.79,443.23){\usebox{\plotpoint}}
\put(1037.91,431.52){\usebox{\plotpoint}}
\put(1054.79,419.44){\usebox{\plotpoint}}
\put(1071.75,407.47){\usebox{\plotpoint}}
\put(1088.62,395.38){\usebox{\plotpoint}}
\put(1105.60,383.46){\usebox{\plotpoint}}
\put(1122.90,372.02){\usebox{\plotpoint}}
\put(1139.38,359.41){\usebox{\plotpoint}}
\put(1156.11,347.17){\usebox{\plotpoint}}
\put(1173.51,335.96){\usebox{\plotpoint}}
\put(1189.71,323.06){\usebox{\plotpoint}}
\put(1206.85,311.40){\usebox{\plotpoint}}
\multiput(1215,305)(16.604,-12.453){2}{\usebox{\plotpoint}}
\put(1256.40,273.70){\usebox{\plotpoint}}
\multiput(1269,264)(16.844,-12.128){2}{\usebox{\plotpoint}}
\put(1306.00,236.09){\usebox{\plotpoint}}
\multiput(1317,227)(17.038,-11.853){2}{\usebox{\plotpoint}}
\put(1355.53,198.43){\usebox{\plotpoint}}
\put(1371.45,185.12){\usebox{\plotpoint}}
\put(1387.45,171.91){\usebox{\plotpoint}}
\put(1403.76,159.07){\usebox{\plotpoint}}
\put(1420.15,146.33){\usebox{\plotpoint}}
\end{picture}
\caption{Dependence of the $\mw$ prediction on the electroweak scale $\mu$
        in the $\ms$ scheme for $\mt=175$ GeV, $\mh=100$
        GeV, including only the leading $O(g^4 \mt^4/\mw^4)$
        correction (dotted curve) or incorporating also the 
        irreducible
        $O(g^4 \mt^2/\mw^2)$ contribution (solid curve). QCD  corrections 
        are not included.}
\end{figure}


\begin{thebibliography}{99}
\bibitem{footnote}
As the errors in $A_3$ are slightly asymmetrical, we identify 
$\sigma_3$ with the larger one. Following Ref. \cite{EWWG}, the lower limit 
on $\mh$ from direct searches is not employed here in the upper bound
 determinations.
\end{thebibliography}
\end{document}